\documentclass[usenatbib]{mn2e}
\usepackage{natbibmnfix,graphicx,times}
\usepackage{amsbsy}

\newcommand{\bq}{\begin{equation}}
\newcommand{\eq}{\end{equation}}
\newcommand{\bqa}{\begin{eqnarray}}
\newcommand{\eqa}{\end{eqnarray}}

\newcommand{\eV}{\mbox{ eV}}
\newcommand{\kel}{\mbox{ K}}
\newcommand{\mkel}{\mbox{ mK}}

\newcommand{\secinv}{\mbox{ s$^{-1}$}}

\newcommand{\xhi}{x_{\rm HI}}

\newcommand{\dtb}{\delta T_b}

\newcommand{\lya}{Ly$\alpha$ }

\newcommand{\deriv}{{\rm d}}

\newcommand{\apj}{ApJ}
\newcommand{\apjl}{ApJ}

\newcommand{\aj}{AJ}
\newcommand{\jgr}{J. Geophys. Research}
\newcommand{\mnras}{MNRAS}

\newcommand{\physrep}{Physics Reports}

\newcommand{\pra}{PRA}

\title[Proton-hydrogen spin exchange rates]{Spin exchange rates in proton-hydrogen collisions}

\author[S.~R. Furlanetto \& M.~R. Furlanetto]{Steven R.  Furlanetto$^1$\thanks{Email:  steven.furlanetto@yale.edu} \& Michael R. Furlanetto $^2$\thanks{Email:  mfurlanetto@lanl.gov}\\
$^1$Yale Center for Astronomy and Astrophysics, Yale University, PO Box 208121, New Haven, CT 06520-8121\\
$^2$Physics Division, Los Alamos National Laboratory, MS H803, P.O. Box 1663, Los Alamos NM 87545}

\begin{document}

\maketitle

\begin{abstract}
The spin temperature of neutral hydrogen, which determines the
optical depth and brightness of the 21 cm line, is determined by the
competition between radiative and collisional processes.  Here we
examine the role of proton-hydrogen collisions in setting the spin
temperature.  We use recent fully quantum-mechanical calculations of
the relevant cross sections, which allow us to present accurate
results over the entire physically relevant temperature range
$1$--$10^4 \kel$.  For kinetic temperatures $T_K \ga 100 \kel$, the
proton-hydrogen rate coefficient exceeds that for hydrogen-hydrogen
collisions by about a factor of two.  However, at low temperatures
($T_K \la 5 \kel$) H--H$^+$ collisions become several thousand times
more efficient than H--H and even more important than H--e$^-$
collisions.
\end{abstract}


\section{Introduction} \label{intro}

The 21 cm transition is potentially a powerful probe of the
pre-reionization intergalactic medium (IGM) because of the enormous
amount of neutral hydrogen in the Universe at that time
\citep{field58, scott90, madau97}.  It can teach us about
reionization, the formation of the first structures and the first
galaxies, and even the ``dark ages" before these objects formed
(\citealt{furl06-review}, and references therein).  It is therefore
crucial to understand the fundamental physics underlying the 21 cm
transition.  One critical aspect is the spin temperature, which is
determined by the competition between the scattering of cosmic microwave
background (CMB) photons, the scattering of \lya photons
\citep{wouthuysen52, field58}, and collisions.  When CMB scattering
dominates, the IGM remains invisible because the spin temperature
approaches that of the CMB (which is used as a backlight).

Before star formation commences, collisions are the only way to
break this degeneracy.  The total coupling rate is determined by
collisions with other hydrogen atoms, protons, and electrons.  At
the low residual electron fraction expected after cosmological
recombination \citep{seager99}, H--H collisions dominate.  Spin
exchange in such interactions has received a great deal of attention
over the years \citep{purcell56, smith66, allison69, zygelman05,
hirata06}.  We have also recently re-examined spin-exchange in
H--e$^-$ collisions using accurate numerical cross-sections
including the $L=0$--$3$ partial waves \citep{furl06-elec} and
showed that such collisions become important when the ionized
fraction $x_i \ga 0.01$.

However, proton-hydrogen collisions have received little attention
in the literature; \citet{smith66} provided the most recent
evaluation of their spin-exchange rate coefficient, but he used only
semi-classical estimates of the cross sections.  In the intervening
years, atomic physicists have calculated the relevant quantum
mechanical cross sections to high accuracy using increasingly
sophisticated numerical techniques (e.g., \citealt{hunter77,
hodges91, krstic99, krstic04}).  Our purpose here will be to
generate similarly accurate rate coefficients from the
\citet{krstic04} cross sections for use in 21 cm calculations.  We
will show that, at sufficiently low temperatures, H--H$^+$
collisions could dominate the spin coupling, but that in more
realistic circumstances they provide only a small correction to the
usual calculation.

The remainder of this paper is organized as follows.  In \S
\ref{21cm}, we briefly review the 21 cm transition.  Our main
results are contained in \S \ref{phcoll}, where we calculate the
spin de-excitation rates for H--H$^+$ collisions.  We conclude and
discuss applications to the 21 cm transition in the high-redshift
IGM in \S \ref{disc}.

\section{The 21 cm Transition} \label{21cm}

We review the relevant characteristics of the 21 cm transition here;
we refer the interested reader to \citet{furl06-review} for a more
comprehensive discussion.  The 21 cm brightness temperature
(relative to the CMB) of a patch of the IGM is
\begin{eqnarray}
\dtb & = & 27 \, \xhi \, (1 + \delta) \, \left( \frac{\Omega_b h^2}{0.023} \right) \left( \frac{0.15}{\Omega_m h^2} \, \frac{1+z}{10} \right)^{1/2} \nonumber \\
& & \times \left( \frac{T_S - T_\gamma}{T_S} \right) \, \left[ \frac{H(z)/(1+z)}{\deriv v_\parallel/\deriv r_\parallel} \right] \mkel,
\label{eq:dtb}
\end{eqnarray}
where $\delta$ is the fractional overdensity, $\xhi$ is
the neutral fraction,
$T_S$ is the spin temperature, $T_\gamma$ is the CMB temperature, and $\deriv
v_\parallel/\deriv r_\parallel$ is the gradient of the proper
velocity along the line of sight.  The last factor accounts for
redshift-space distortions  \citep{bharadwaj04-vel,barkana05-vel}.

The spin temperature $T_S$ is determined by competition between
scattering of CMB photons, scattering of UV photons
\citep{wouthuysen52, field58}, and collisions \citep{purcell56}.  In
equilibrium,
\begin{equation}
T_S^{-1} = \frac{T_\gamma^{-1} + x_c T_K^{-1} + x_\alpha T_c^{-1}}{1 + x_{c} + x_\alpha},
\label{eq:tsdefn}
\end{equation}
where $T_K$ is the kinetic temperature of the gas, $T_c$ is the
color temperature of the radiation field at the \lya transition, and
the $x_i$ are coupling coefficients. The last part of this equation
describes the Wouthuysen-Field effect, in which the absorption and
re-emission of \lya photons mixes the hyperfine states
\citep{wouthuysen52, field58}.  We refer the reader to
\citet{furl06-review}, and references therein, for more detail on
this component.

The factor $x_c$ is the total collisional coupling coefficient, including H--H,
H--e$^-$, and H--H$^+$ collisions.  In this paper, we will focus
only on the contribution from proton-hydrogen collisions , which we
will denote $x_c^{\rm pH}$; the other components are $x_c^{\rm HH}$
and $x_c^{\rm eH}$, with obvious meanings (see \citealt{zygelman05}
and \citealt{furl06-elec} for the most recent estimates).  In more
detail, \bq x_c^{\rm pH} = \frac{n_p \kappa_{10}^{\rm pH}
T_\star}{A_{10} T_\gamma}, \label{eq:xc} \eq where $\kappa_{10}^{\rm
pH}$ is the spin de-excitation rate in proton-hydrogen collisions,
$n_p$ is the local proton density, $T_\star \equiv h \nu_{21}/k_B =
0.068 \kel$, $\nu_{21}$ is the frequency of the 21 cm line, and
$A_{10} = 2.85 \times 10^{-15} \secinv$ is the Einstein-$A$
coefficient for that transition.

\section{Proton-Hydrogen Collisions} \label{phcoll}

\subsection{The spin exchange cross section} \label{cross}

We must first compute the cross section for spin exchange as a
function of collision energy.  To do so, we begin by noting that the
radial wave functions in the H--H$^+$ system satisfy the Schr{\"
o}dinger equations in uncoupled partial waves of angular momentum
$L$.  In the energy range with which we are concerned, these become
\bq \left[ \frac{\deriv^2}{\deriv R^2} - \frac{L (L + 1)}{R^2} - 2
\mu E_i(R) + 2 \mu E \right] \Psi_i^{(L)}(R) = 0, \label{eq:schrod}
\eq where $R$ is the internuclear distance, $\mu$ is the reduced
mass, $E_i$ is the adiabatic electronic potential for the $1s
\sigma_g$ (gerade) or $2 p \sigma_u$ (ungerade) states of H$_2^+$,
and $\Psi_i$ is the radial wave function for the appropriate
channel.

We assume the process to be electronically elastic, so that the
scattering problem can be reduced in the usual way to the
computation of phase shifts $\delta_L^g$ and $\delta_L^u$ (see
\citealt{furl06-elec} for a description of the analogous
transformation for H--e$^-$ scattering).  The so-called charge
transfer cross section $\sigma_{\rm ct}$ may then be written (e.g.,
\citealt{krstic04}) \bq \sigma_{\rm ct} = \frac{4 \pi}{k^2}
\sum_{L=0}^{\infty} (2 L + 1) \sin^2 (\delta_L^g - \delta_L^u),
\label{eq:sigmact} \eq where $k$ is the center-of-mass momentum.
This cross section is defined with reference to an experiment in
which a beam of polarized protons (with, say, spin +1/2) is incident
on an unpolarized collection of hydrogen atoms and in which the
spins of the scattered protons are measured, so that protons with
spin --1/2 can be unambiguously determined to have originated in the
target. It tends to the usual charge transfer cross section in the
classically distinguishable particle limit \citep{krstic99-defn}.

This cross section must be calculated numerically, and there is a
long history of such attempts \citep{hunter77, hodges91, krstic99}.
For $10^{-4} \eV < E < 10^2 \eV$, we use the most recent
calculations, from \citealt{krstic04}, which are accurate numerically to six significant figures, although
the neglect of coupling to higher electronic states reduces the
accuracy somewhat at the upper end of the energy range.

\begin{figure}
\begin{center}
\resizebox{8cm}{!}{\includegraphics{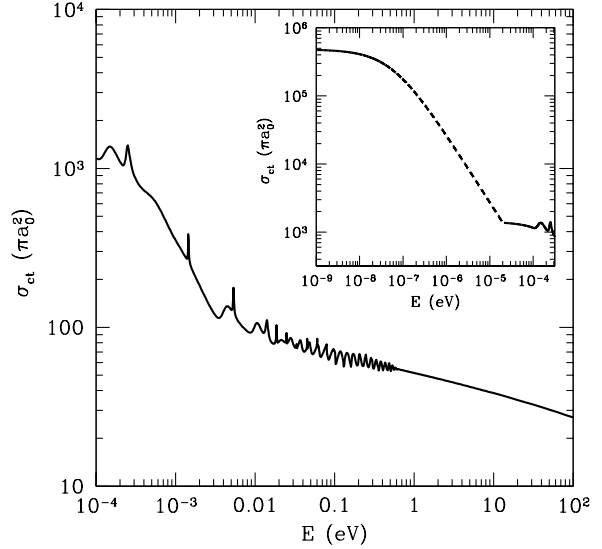}}\\%
\end{center}
\caption{Charge transfer cross section (in units of $\pi a_0^2$) for H--H$^+$ collisions, as a function of the collision energy.  The solid curve shows $\sigma_{\rm ct}$ from \citet{krstic04}.  The dashed curve in the inset shows the effective range approximation for the low energy behavior (see text).  } \label{fig:cs}
\end{figure}

Figure~\ref{fig:cs} shows the resulting cross section, in units of
$\pi a_0^2$ (where $a_0$ is the Bohr radius) over this entire energy
range.  Note the rich structure in $\sigma_{\rm ct}$, especially at
moderately small energies.  This is in sharp contrast to the
H--e$^-$ cross section, which is smooth until the neighborhood of
the $n=2$ excitation threshold (see Fig. 1 in
\citealt{furl06-elec}).  Capturing this detailed structure required
the use of a closely-spaced energy grid (664 points were used here)
and the inclusion of high-$L$ partial waves (up to $L_{\rm
max}=3200$ at 100 eV; \citealt{krstic04}).  The narrow features are
caused by shape resonances in the electronic potentials and have the
usual Fano line shape. The broader features, known as ``Regge
oscillations," have a more subtle origin.  They are generated by combinations of one to three poles in the
$S$-matrix, each of which corresponds to one of the $L=0$
vibrational bound states of H$_2^+$.  These features are difficult
to understand in the traditional partial wave representation, but
their nature has recently been explained using the Mulholland
representation \citep{macek04}. Note that the charge transfer cross
section shows significantly less structure than the elastic cross
section, which also includes broad oscillations at higher energies
due to the glory effect \citep{child84}.

\begin{figure*}
\begin{center}
\resizebox{8cm}{!}{\includegraphics{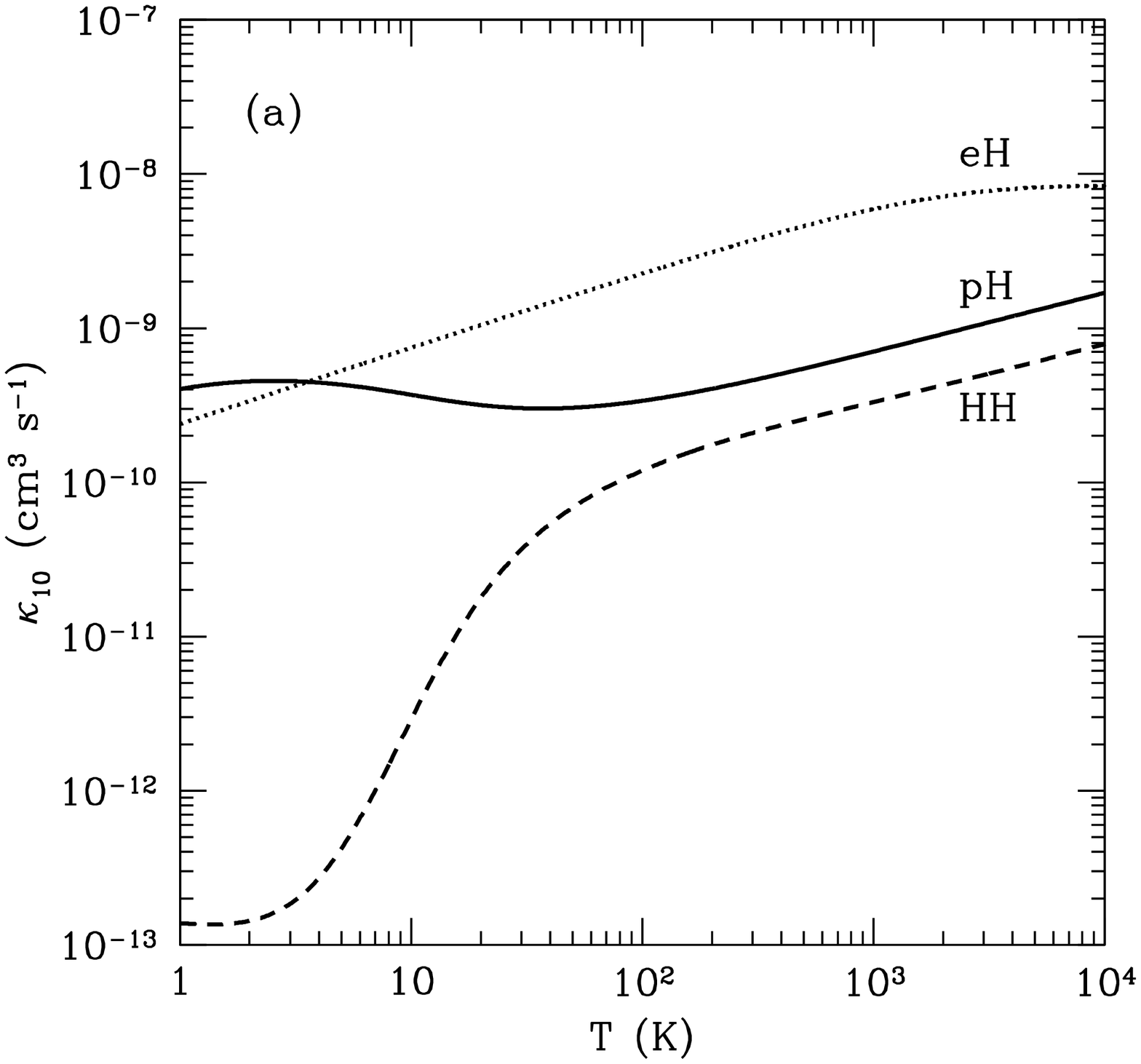}}
\hspace{0.13cm}
\resizebox{8cm}{!}{\includegraphics{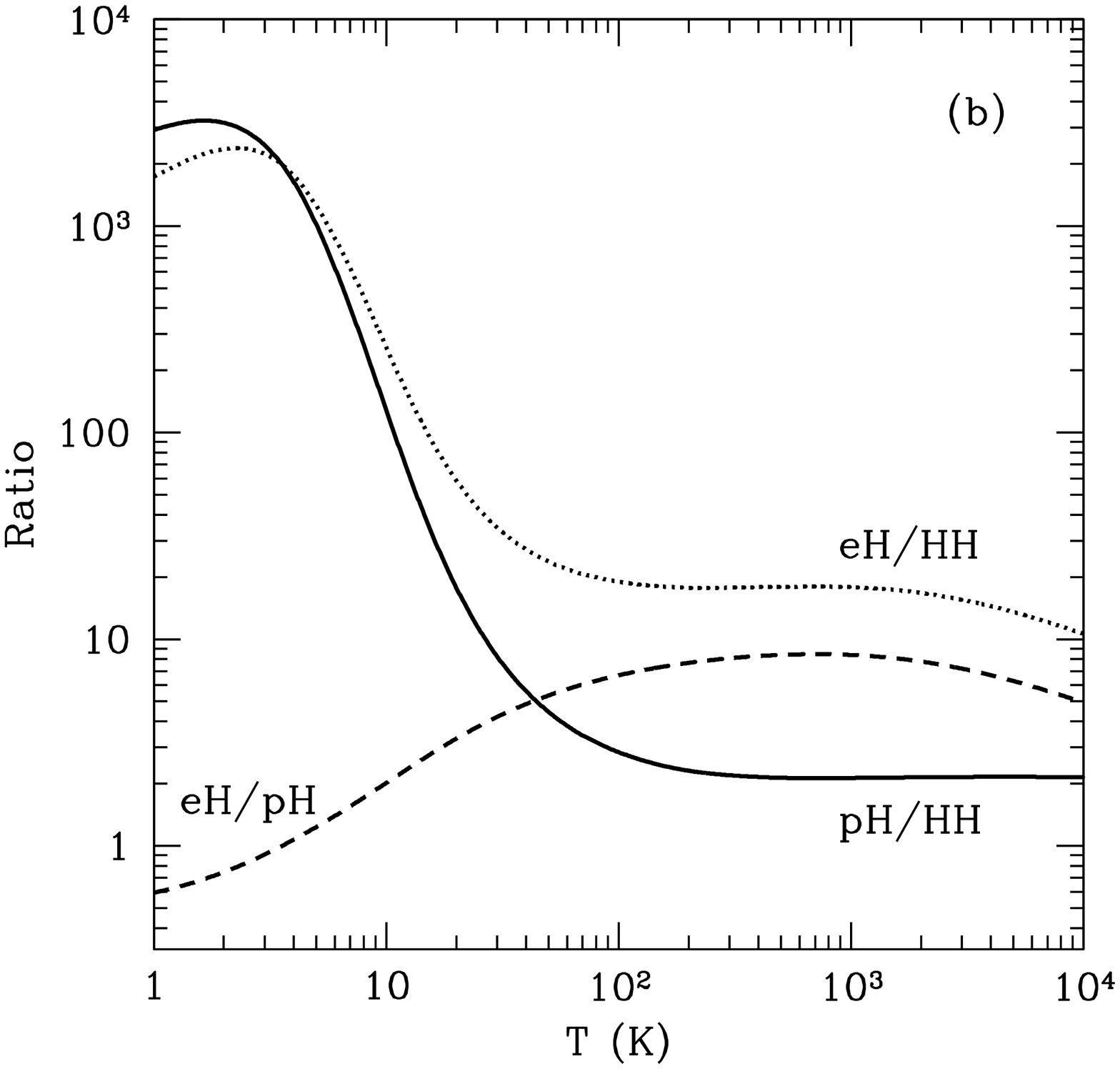}}
\end{center}
\caption{Spin-exchange rate coefficients.  \emph{(a)}:  Rate coefficients for H--H$^+$ collisions (solid curve), H--e$^+$ collisions (dotted curve, from \citealt{furl06-elec}), and H--H collisions (dashed curve, from \citet{zygelman05}).  \emph{(b)}:  Ratios between the three rate coefficients in panel \emph{a}.} \label{fig:kappa}
\end{figure*}

These calculations ignored non-adiabatic and relativistic effects,
so they were not extended below $\sim 10^{-4} \eV$, where such
phenomena become significant. Instead, we use the numerically
converged low-energy cross sections of \citet{glassgold05}. These
calculations used the same technique as \citet{krstic04}, and thus
also neglected non-adiabatic and relativistic effects, but they at
least provide us with a baseline estimate of the collision behavior
at small energies. They show that only the $L=0$ terms remain
non-zero at low energies, and $\delta_0^g \rightarrow n\pi$ as $E
\rightarrow 0$. In this limit, \bq \sigma_{\rm ct} \approx \frac{4
\pi}{k^2} \sin^2 \delta_0^u, \label{eq:sigma_e0} \eq which is
one-quarter of the total elastic cross section.  Meanwhile,
effective-range theory yields an excellent approximation for
$\delta_0^u$: \bq k \cot \delta_0^u = - \frac{1}{a} + \frac{\pi
\alpha}{3 a^2} k + \frac{2 \alpha}{3 a} k^2 \ln \left( \frac{\alpha
k^2}{16} \right) + O(k^3), \label{eq:delta0u} \eq where the
scattering length $a=801.2$ $a_0$ and the polarizability
$\alpha=4.5$  $a_0^3$ parameterize the polarization potential of the
hydrogen atom \citep{glassgold05}.  

The resulting cross section is shown for $10^{-10} \eV < E < 10^{-4}
\eV$ by the inset in Figure~\ref{fig:cs}.  We use the
effective-range approximation for $E<2 \times 10^{-5} \eV$ and
linearly interpolate the cross section between that value and
$10^{-4} \eV$ (where \citealt{glassgold05} showed that the effective
range approximation for the total cross section breaks down).  This effective range theory approximation provides a good match to numerical calculations of the spin-exchange cross section (using methods similar to \citealt{glassgold05}) in this energy regime (P. Krsti{\' c}, private communication).
Although we do not know how strongly the relativistic and
non-adiabatic corrections will affect the cross sections, we do not
expect them to have a dramatic effects on our final results, as
discussed below.

\subsection{The spin-exchange rate coefficient}
\label{rate}

To apply this to H--H$^+$ collisions in the IGM, we must thermally
average the above cross sections.  The spin-exchange rate
coefficient relevant for our purposes is \bq \kappa_{10}^{\rm pH} =
\sqrt{ \frac{8 k_B T_K}{\pi \mu} } \left( \frac{3}{16} \sigma_{\rm
ct} \right), \label{eq:kappa} \eq where $k_B$ is Boltzmann's
constant, $T_K$ is the kinetic temperature of the IGM, and the
prefactor with the square root is the thermally-averaged velocity.
The factor $3/16$ appears because the cross section as defined above
assumes a polarized beam of electrons, while here we care about the
net rate at which spin de-excitation occurs \citep{smith66,
krstic99-defn}.  We must therefore multiply $\sigma_{\rm ct}$ by
$1/4$ (the probability that the target atom has $F=0$) and then
$3/4$ (the probability that the scattered atom has $F=1$).

The solid line in Figure~\ref{fig:kappa}\emph{a} shows
$\kappa_{10}^{\rm pH}$ over the full temperature range of interest.
It is much smoother than $\sigma_{\rm ct}$, because the thermal
averaging washes out the resonances.  At moderate to high
temperatures, the rate coefficient is roughly proportional to
$T_K^{1/2}$, because $\sigma_{\rm ct}$ is itself falling only slowly
with energy.  However, at lower temperatures the cross section
increases rapidly, so that $\kappa_{10}^{\rm pH}$ is actually the
largest rate coefficient at $T_K \la 3 \kel$.  This is in sharp
contrast to hydrogen-hydrogen and hydrogen-electron collisions,
whose rate coefficients are shown by the dashed and dotted lines,
respectively. The H--e$^-$ cross section increases as $E \rightarrow
0$, but only modestly (see Fig. 1 of \citealt{furl06-elec}), while
the H--H rate decreases rapidly as $E \rightarrow 0$ (see below).

Figure~\ref{fig:kappa}\emph{b} compares these three processes in
more detail, showing the ratios between the various cross sections.
At extremely small temperatures, H--H$^+$ collisions can be the
dominant process.  Electron-hydrogen collisions quickly come to dominate at $T_K \ga 3
\kel$, and remain more important at all higher temperatures.
Naively, we would expect that $\kappa_{10}^{\rm eH}/\kappa_{10}^{\rm
pH} \sim \sqrt{m_H/2 m_e} \sim 30$.  In reality, the ratio is
several times smaller than that, because proton-hydrogen collisions
have a substantially larger cross section throughout this
temperature range.\footnote{Note that the decline in
$\kappa_{10}^{\rm eH}/\kappa_{10}^{\rm pH}$ at $T \ga 3000 \kel$ is
somewhat artificial.  \citet{furl06-elec} excluded all electrons
with $E>10.2 \eV$ from the thermal average, because these electrons
have a substantial probability of exciting the atom to higher electronic
states.}

Figure~\ref{fig:kappa}\emph{b} also shows that, at high
temperatures, H--H$^+$ collisions are marginally more efficient than
H--H collisions (by about a factor of two); the shapes of these two
rate coefficients mirror each other quite closely in this regime.
However, at lower temperatures, proton-hydrogen collisions are
vastly more efficient. As we have seen, the $L=0$ term in
$\sigma_{\rm ct}$ approaches a large constant value; in the
analogous H--H problem, the corresponding quantity actually
approaches zero because of an accidental cancellation in the
$S$-wave cross sections \citep{zygelman05,sigurdson05-deut}

For use in other calculations, we present in Table~\ref{tab:kappa}
our results for $\kappa_{10}^{\rm pH}$ as a function of temperature.
These are fully numerically converged to the six significant figure
accuracy of the \citet{krstic04} cross sections, but as emphasized
above they do not include all the relevant physical processes across
the entire temperature range.  In particular, at the lower end of
the temperature range, non-adiabatic processes and relativistic
corrections lead to uncertainties at the per cent level.  For
example, assuming that the cross section is constant below $10^{-4}
\eV$ changes our final results by $(5,\,1,\,0.2)$ per cent at
$T_K=(1,\,3.7,\,10) \kel$, respectively.  At the upper end of the
temperature range, rotational coupling of the $2p \sigma_u$ and $2p
\pi_u$ states affects the rate coefficients at the $\sim 0.1$ per
cent level.

\begin{table}
\begin{center}
\caption{Proton-hydrogen spin de-excitation rate coefficients}
\label{tab:kappa}
\begin{tabular}{|c|c|c|c|}
\hline
$T_K \, ({\rm K})$ & $\kappa_{10}^{\rm pH} \, (10^{-9} \, {\rm cm^3 \ s^{-1}})$ &$T_K \, ({\rm K})$ & $\kappa_{10}^{\rm pH} \, (10^{-9} \, {\rm cm^3 \ s^{-1}})$  \\
\hline
1 & 0.4028 & 1000 & 0.7051 \\
2 & 0.4517 & 2000 & 0.9167 \\
5 & 0.4301 & 3000 & 1.070 \\
10 & 0.3699 & 5000 &  1.301 \\
20 & 0.3172 & 7000 & 1.480 \\
50 & 0.3047 & 10,000 & 1.695 \\
100 & 0.3379 & 15,000 & 1.975 \\
200 & 0.4043 & 20,000 & 2.201 \\
500 & 0.5471 & & \\
\hline
\end{tabular}
\end{center}
\end{table}

\section{Discussion} \label{disc}

We have computed the rate coefficients for spin-exchange in H--H$^+$
collisions using recent, fully quantum mechanical solutions for the
relevant cross sections over most of the energy range of interest
\citep{krstic04}.  We have also extended the calculation to lower
temperatures ($T_K \sim 1 \kel$) by using the approximate cross
sections of \citet{glassgold05}; while these ignore non-adiabatic
effects and relativistic corrections, they appear to be accurate to
several per cent.

Our results (collected in numerical form in Table~\ref{tab:kappa})
will be particularly useful for calculating the spin temperature of
the high-redshift IGM before reionization (and hence brightness in
the 21 cm line).  Before the first luminous sources appear, this is
determined purely by the competition between collisions and CMB
scattering.  In most cases, protons contribute a small but
non-negligible fraction of the coupling.  For example, the standard
recombination calculation yields a global ionized fraction $x_i \sim
2 \times 10^{-4}$ and $T_K \sim 9 [(1+z)/20]^{1.85} \kel$ for
redshifts $z \sim 10$--$100$ \citep{seager99}; the exponent is
slightly smaller than two (which would be expected for an
adiabatically-cooling non-relativistic gas) because of a small
amount of Compton heating off the CMB.  At the higher redshifts,
where $T_K \sim 100 \kel$, protons account for only $\sim 0.04$ per
cent of the total coupling, but if the gas does remain cool to $z
\sim 20$ they provide $\sim 2$ per cent as much coupling as
hydrogen-hydrogen collisions.  In the unlikely event that the gas
cools to $T=2.5 \kel$ at $z=10$ without interference from luminous
sources, we would have $x_c^{\rm pH}/x_c^{\rm HH} \sim 0.6$ -- and
protons would become even more important than electrons.

In reality, the first galaxies probably appear at $z \ga 20$.  They
flood the Universe with \lya photons (which affect the spin
temperature through the Wouthuysen-Field mechanism) and X-ray
photons (which heat the gas).  As a result, the spin temperature
probably increases well before $z=10$ \citep{sethi05, furl06-glob,
pritchard06}, so that protons never actually come to dominate the
coupling.  However, they still must be included for high-accuracy
predictions of the 21 cm signal.

In this paper, we have only examined spin-exchange via
electronically elastic proton-hydrogen collisions. At high
temperatures, such collisions can instead excite higher atomic
levels or even ionize the atom.  Such processes could also affect
the spin temperature because the hydrogen atom could enter a
different spin state after the radiative cascade that follows.
Although we do not know the relevant excitation cross sections, our
experience with electron-hydrogen collisions suggests that in
practice these collisions will not be important. For H--e$^-$, the
most likely transition at relatively low temperatures is excitation
to the $2p$ state, which is followed almost immediately by radiative
de-excitation and emission of a \lya photon.  This \lya photon
scatters $\sim 10^5$ times before redshifting out of resonance, so
in terms of spin coupling it is much more important than the single
collision that generated it. For a Maxwell-Boltzmann energy
distribution, excitation of this process through electron-hydrogen
collisions suddenly comes to dominate at $T_K \sim 6300 \kel$
\citep{furl06-elec}. If proton-hydrogen collisions cause similar
excitations, they too would become relatively unimportant once
electronic excitations become energetically feasible. Furthermore,
because the electron-hydrogen rate coefficient exceeds the
proton-hydrogen rate coefficient by a factor of several in this
temperature range, we expect proton collisions to be only a minor
perturbation. At higher temperatures, the collisionally-generated
\lya photons completely dominate, and the effect of
$\kappa_{10}^{\rm pH}$ will be even smaller.\footnote{In reality,
X-ray heating of the IGM results in an excess of fast electrons even
when the temperature is much smaller. Thus the \lya channel can be
important throughout the ``reheating" era \citep{chen06,
chuzhoy06-first, pritchard06}.}

\vspace{0.1cm}

We thank P. Krsti{\' c} for making his H--H$^+$ cross section data
available to us in an electronic form.  This publication has been
approved for release as LA-UR-07-0513.  Los Alamos National
Laboratory, an affirmative action/equal opportunity employer, is
operated by Los Alamos National Security, LLC, for the National
Nuclear Security Administration of the U.S. under contract
DE-AC52-06NA25396.


\begin{thebibliography}{}

\bibitem[\protect\citeauthoryear{{Allison} \& {Dalgarno}}{{Allison} \&
  {Dalgarno}}{1969}]{allison69}
{Allison} A.~C.,  {Dalgarno} A.,  1969, \apj, 158, 423

\bibitem[\protect\citeauthoryear{{Barkana} \& {Loeb}}{{Barkana} \&
  {Loeb}}{2005}]{barkana05-vel}
{Barkana} R.,  {Loeb} A.,  2005, \apjl, 624, L65

\bibitem[\protect\citeauthoryear{{Bharadwaj} \& {Ali}}{{Bharadwaj} \&
  {Ali}}{2004}]{bharadwaj04-vel}
{Bharadwaj} S.,  {Ali} S.~S.,  2004, \mnras, 352, 142

\bibitem[\protect\citeauthoryear{{Chen} \& {Miralda-Escude}}{{Chen} \&
  {Miralda-Escude}}{2006}]{chen06}
{Chen} X.,  {Miralda-Escude} J.,  2006, submitted to \apj \ (astro-ph/0605439)

\bibitem[\protect\citeauthoryear{{Child}}{{Child}}{1984}]{child84}
Child M.~S., 1984, Molecular Collision Theory (Dover Publications, Inc.:  Mineola, NY)

\bibitem[\protect\citeauthoryear{{Chuzhoy}, {Alvarez} \& {Shapiro}}{{Chuzhoy}
  et~al.}{2006}]{chuzhoy06-first}
{Chuzhoy} L.,  {Alvarez} M.~A.,    {Shapiro} P.~R.,  2006, \apjl, 648, L1

\bibitem[\protect\citeauthoryear{{Field}}{{Field}}{1958}]{field58}
{Field} G.~B.,  1958, Proc. I.R.E., 46, 240

\bibitem[\protect\citeauthoryear{{Furlanetto}}{{Furlanetto}}{2006}]{furl06-glob}
{Furlanetto} S.~R.,  2006, \mnras, 371, 867

\bibitem[\protect\citeauthoryear{{Furlanetto} \& {Furlanetto}}{{Furlanetto} \&
  {Furlanetto}}{2007}]{furl06-elec}
{Furlanetto} S.~R.,  {Furlanetto} M.~R.,  200, \mnras, 374, 547

\bibitem[\protect\citeauthoryear{{Furlanetto}, {Oh} \& {Briggs}}{{Furlanetto}
  et~al.}{2006}]{furl06-review}
{Furlanetto} S.~R.,  {Oh} S.~P.,    {Briggs} F.~H.,  2006, \physrep, 433, 181

\bibitem[\protect\citeauthoryear{{Glassgold}, {Krsti{\'c}} \&
  {Schultz}}{{Glassgold} et~al.}{2005}]{glassgold05}
{Glassgold} A.~E.,  {Krsti{\'c}} P.~S.,    {Schultz} D.~R.,  2005, \apj, 621,
  808

\bibitem[\protect\citeauthoryear{{Hirata} \& {Sigurdson}}{{Hirata} \&
  {Sigurdson}}{2006}]{hirata06}
{Hirata} C.~M.,  {Sigurdson} K.,  2006, submitted to \mnras \
  (astro-ph/0605071)

\bibitem[\protect\citeauthoryear{{Hodges} Jr. \& {Breig}}{{Hodges} \&
  {Breig}}{1991}]{hodges91}
{Hodges} Jr. R.~R.,  {Breig} E.~L.,  1991, \jgr, 96, 7697

\bibitem[\protect\citeauthoryear{{Hunter} \& {Kuriyan}}{{Hunter} \&
  {Kuriyan}}{1977}]{hunter77}
{Hunter} G.,  {Kuriyan} M.,  1977, Royal Society of London Proceedings Series
  A, 353, 575

\bibitem[\protect\citeauthoryear{{Krsti{\'c}}, {Macek}, {Ovchinnikov} \&
  {Schultz}}{{Krsti{\'c}} et~al.}{2004}]{krstic04}
{Krsti{\'c}} P.~S.,  {Macek} J.~H.,  {Ovchinnikov} S.~Y.,    {Schultz} D.~R.,
  2004, \pra, 70, 042711

\bibitem[\protect\citeauthoryear{{Krsti{\'c}} \& {Schultz}}{{Krsti{\'c}} \&
  {Schultz}}{1999a}]{krstic99-defn}
{Krsti{\'c}} P.~S.,  {Schultz} D.~R.,  1999a, \pra, 60, 2118

\bibitem[\protect\citeauthoryear{{Krsti{\'c}} \& {Schultz}}{{Krsti{\'c}} \&
  {Schultz}}{1999b}]{krstic99}
{Krsti{\'c}} P.~S.,  {Schultz} D.~R.,  1999b, Journal of Physics B Atomic
  Molecular Physics, 32, 3485

\bibitem[\protect\citeauthoryear{{Macek}, {Krsti{\'c}} \&
  {Ovchinnikov}}{{Macek} et~al.}{2004}]{macek04}
{Macek} J.~H.,  {Krsti{\'c}} P.~S.,    {Ovchinnikov} S.~Y.,  2004, Physical
  Review Letters, 93, 183203

\bibitem[\protect\citeauthoryear{{Madau}, {Meiksin} \& {Rees}}{{Madau}
  et~al.}{1997}]{madau97}
{Madau} P.,  {Meiksin} A.,    {Rees} M.~J.,  1997, \apj, 475, 429

\bibitem[\protect\citeauthoryear{{Pritchard} \& {Furlanetto}}{{Pritchard} \&
  {Furlanetto}}{2006}]{pritchard06}
{Pritchard} J.~R.,  {Furlanetto} S.~R.,  2006, submitted to \mnras \
  (astro-ph/0607234)

\bibitem[\protect\citeauthoryear{{Purcell} \& {Field}}{{Purcell} \&
  {Field}}{1956}]{purcell56}
{Purcell} E.~M.,  {Field} G.~B.,  1956, \apj, 124, 542

\bibitem[\protect\citeauthoryear{{Scott} \& {Rees}}{{Scott} \&
  {Rees}}{1990}]{scott90}
{Scott} D.,  {Rees} M.~J.,  1990, \mnras, 247, 510

\bibitem[\protect\citeauthoryear{{Seager}, {Sasselov} \& {Scott}}{{Seager}
  et~al.}{1999}]{seager99}
{Seager} S.,  {Sasselov} D.~D.,    {Scott} D.,  1999, \apjl, 523, L1

\bibitem[\protect\citeauthoryear{{Sethi}}{{Sethi}}{2005}]{sethi05}
{Sethi} S.~K.,  2005, \mnras, 363, 818

\bibitem[\protect\citeauthoryear{{Sigurdson} \& {Furlanetto}}{{Sigurdson} \&
  {Furlanetto}}{2006}]{sigurdson05-deut}
{Sigurdson} K.,  {Furlanetto} S.~R.,  2006, Physical Review Letters, 97, 091301

\bibitem[\protect\citeauthoryear{{Smith}}{{Smith}}{1966}]{smith66}
{Smith} F.~J.,  1966, Plan. Space Sci., 14, 929

\bibitem[\protect\citeauthoryear{{Wouthuysen}}{{Wouthuysen}}{1952}]{wouthuysen%
52}
{Wouthuysen} S.~A.,  1952, \aj, 57, 31

\bibitem[\protect\citeauthoryear{{Zygelman}}{{Zygelman}}{2005}]{zygelman05}
{Zygelman} B.,  2005, \apj, 622, 1356

\end{thebibliography}

\end{document}